\documentstyle[12pt,leqno]{article}

\newcommand{\beq}{\begin{equation}}
\newcommand{\eeq}{\end{equation}}
\newcommand{\bear}{\begin{array}}
\newcommand{\ear}{\end{array}}

\begin{document}

\vspace {4em}

{\flushright hep-th/9303092}

\vspace {4em}
Proc. NATO ARW "Applications of analitic and geometric methods
   to nonlinear differential equations, 14-19 July 1992, Exeter, UK)

\vspace {4em}

\noindent {\bf CLASSICAL DIFFERENTIAL GEOMETRY AND INTEGRABILITY OF
  SYSTEMS OF HYDRODYNAMIC TYPE }
\vspace{3em}

\noindent\hspace{9ex}SERGUEI P. TSAREV

\noindent\hspace{9ex}{\it Steklov Mathematical Institute}

\noindent\hspace{9ex}{\it 117966, Vavilova,42,}

\noindent\hspace{9ex}{\it Moscow, GSP-1, USSR}

\noindent\hspace{9ex}e-mail: tsarev@top.mian.su

\vspace{1em}
\noindent\hspace{9em}August 25, 1992

\vspace{2em}

\noindent ABSTRACT. Remarkable parallelism between the theory of
integrable systems of first-order quasilinear PDE and some old results in
projective and affine differential geometry of conjugate nets, Laplace
equations, their Bianchi-B\"acklund transformations is exposed.
 These results were recently
applied by I.M.Krichever and B.A.Dubrovin to prove integrability of some
models in topological field theories. Within the geometric framework we
derive some new integrable (in a sense to be discussed) generalizations
describing $N$-wave resonant interactions.

\vspace{2em}

\noindent Ten years ago [10] a natural hamiltonian formalism was proposed
for the class of homogeneous systems of PDE
\vspace{1em}

\noindent (1+1,h)  \qquad \qquad
  $ \bear{l}\left(\bear{c}u^1_t \\ \vdots \\ u^n_t \ear \right)=
   \left(\bear{ccc}v^1_1(u) & \cdots & v^1_n(u) \\
  \cdot & \cdots & \cdot\\
 v^n_1(u) & \cdots & v^n_n(u)  \ear \right)
   \left(\bear{c}u^1_x \\ \vdots \\ u^n_x \ear \right),\\
 u^i=u^i(x,t), \quad i=1, \cdots ,n \ear $

(called "one-dimensional systems of hydrodynamic type"). Later (see [11]) it
was generalized for the class of multidimensional
\vspace{1em}

\noindent (N+1,h) \quad
  $\bear{l} \left(\bear{c}u^1_t \\ \vdots \\ u^n_t \ear \right)=
   \left(\bear{ccc}v^{(1)1}_1(u) & \cdots & v^{(1)1}_n(u) \\
  \cdot & \cdots & \cdot\\
 v^{(1)n}_1(u) & \cdots & v^{(1)n}_n(u)  \ear \right)
   \left(\bear{c}u^1_{x_1} \\ \vdots \\ u^n_{x_1} \ear \right)+ \ldots
\\
   \ldots + \left(\bear{ccc}v^{(N)1}_1(u) & \cdots & v^{(N)1}_n(u) \\
  \cdot & \cdots & \cdot \\
 v^{(N)n}_1(u) & \cdots & v^{(N)n}_n(u)  \ear \right)
   \left(\bear{c}u^1_{x_N} \\ \vdots \\ u^n_{x_N} \ear \right),
\\ u^i=u^i(t,x_1, \ldots ,x_N),
\quad i=1, \ldots ,n \ear $

and non-homogeneous
\vspace{1em}

\noindent (1+1,nh)  \qquad \qquad
  $ \bear{l}\left(\bear{c}u^1_t \\ \vdots \\ u^n_t \ear \right)=
   \left(\bear{ccc}v^1_1(u) & \cdots & v^1_n(u) \\
  \cdot & \cdots & \cdot\\
 v^n_1(u) & \cdots & v^n_n(u)  \ear \right)
   \left(\bear{c}u^1_x \\ \vdots \\ u^n_x \ear \right)+
   \left(\bear{c}f_1(u) \\ \vdots \\ f_n(u) \ear \right),\\
 u^i=u^i(x,t), \quad i=1, \ldots ,n \ear $

\noindent systems.

Some systems (1+1,h) of physical importance such as
 Whitham equations (the averaged 1-phase KdV equation) and Benney equations
have the notable property of being
diagonalizable: under a suitable choice of field variables $u^i$ (Riemann
invariants) the equations become
\begin{equation}
     u^{i}_{t}(x) = v_{i}(u)u^{i}_{x}
\end{equation}
(there is no summation over $i$!). As we have proved in [32], these properties
(hamiltonian property and diagonalizability) imply integrability. Deeper
insight into this type of integrability is given by the theory of
orthogonal curvilinear coordinate systems. This chapter of classical
differential geometry was being intensively developed at the beginning of
the XX century ([6], [8], [20]). In fact this theory gives the geometric
background for integrability of systems (1+1,h), (N+1,h), (1+1,nh).
These forgotten corners of differential geometry seem to be worth revisiting.

{\it An example}. The well-known Bullough-Dodd-Jiber-Shabat equation
$u_{xx}-u_{tt}=e^u - e^{-2u}$  (in the form $(\ln\,h)_{uv}=h-1/h^2 $) was
introduced for the first time in [35] where the respective linear problem
was given as well as a proper B\"acklund transformation for it! It is much
simpler and "geometric" than B\"acklund transformations discussed recently
[1], [31] in the context of integrable systems.

  In this paper we will sketch some applications of methods originating
from classical differential geometry to equations of types (1+1,h),
(N+1,h), (1+1,nh).
\vspace{2em}

\noindent {\bf 1. Diagonal systems of hydrodynamic type and orthogonal
curvilinear coordinate systems in $R^n$}

\vspace{2em}
\noindent Let us recall briefly the main results of [10],[32]. A
(generally nondiagonal) system  $u^{i}_{t} =
\sum^n_{j=1}v^{i}_{j}(u)u^{j}_{x}$  is hamiltonian if there exist a
hamiltonian $H = \int h(u)\,dx$ and a hamiltonian operator $$ \hat{A}^{ij}
= g^{ij}(u)\,\frac{d}{dx} + b_{k}^{ij}(u)\,u_{x}^{k} $$ which define a
skew-symmetric Poisson bracket on functionals
$$
\{I,J\} = \int \frac{\delta I}{\delta  u^{i}(x)}\hat{A}_{ij}
 \frac{\delta J}{\delta u^{j}(x)}dx
$$
satisfying the Jacobi identity  and generate the system
\begin{equation}
     u^{i}_{t}(x) = \{u^{i}(x),H\} =
   \hat{A}_{ij}\frac{\partial H}{\partial u^{j}(x)}  =
  (g^{ij}\partial _{k} \partial _{j} h + b^{ij}_{k}\partial _{j} h)u_{x}^{k}
  = v^{i}_{k} (u)u^{k}_{x}
\end{equation}
where $\partial _{s}=\partial /\partial u^{s}$. B.A.Dubrovin and
S.P.Novikov [10] proved that the necessary and sufficient conditions for
$\hat{A}_{ij}$ to be a hamiltonian operator in the case of non-degeneracy of
the
matrix $g^{ij}$ are:
\begin{quote}
   a)  $g^{ij} = g^{ji}$, i.e. the inverse matrix $g^{-1}$ defines a
Riemannian metric.
\end{quote}
\begin{quote}
   b)    $b^{ij}_{k} = -g^{is} \Gamma ^{j}_{sk}$ for the standard
Christoffel symbols $\Gamma ^{j}_{sk}$ generated by $g_{ij}$.
\end{quote}
\begin{quote}
   c)   the metric $g_{ij}$ has identically vanishing curvature tensor.
\end{quote}

In such case we have  $v^{i}_{j}(u) = \nabla ^{i}\nabla _{j}h =
g^{is}\nabla _{s}\nabla _{j}h$ with the covariant derivatives defined by
$g_{ij}$.

 \underline{Lemma} [32], [33]. In order that
a matrix $v^{i}_{j}(u)$ be a matrix of a
hamiltonian system (1+1,h) with a nondegenerate metric in $ \hat{A}^{ij}$
it is necessary and sufficient that there exists a nondegenerate zero
curvature  metric $g_{ij}$ such that

a) $g_{ik}v^{k}_{j}  = g_{jk}v^{k}_{i} $ and

b) $ \nabla _{j}v^{i}_{k} = \nabla _{k}v^{i}_{j}$,
 where $\nabla $ is the covariant differentiation generated by the
metric $g_{ij}$.

For a {\em diagonal} matrix
$v^{i}_{j}(u) = v_{j}(u)\delta ^{i}_{j}$ this implies that (see [32], [33])
  $g_{ij}$ is also {\it diagonal} and
\beq
 \frac{\partial_{i}v_{k}}{(v_{i}-v_{k})}=\Gamma^{k}_{ki}=
    \frac{1}{2}\partial_{i}\ln\,g_{kk}, \quad
 \partial_{i}=\partial /\partial u^{i}
\eeq
 (hereafter we do not imply the summation on repeated indices!).
{}From (3) we deduce
\begin{equation}
  \partial_{j} \frac{\partial_{i}v_{k}}{v_{i}-v_{k}}=
     \partial_{i} \frac{\partial_{j}v_{k}}{v_{j}-v_{k}}
 , \;\;  i\neq j\neq k.
\end{equation}

  From a differential geometric point of view, to give a zero curvature
nondegenerate diagonal metric is equivalent to giving an orthogonal
curvilinear coordinate system on a flat (possibly pseudo-Euclidean) space
(see [6]). Locally these coordinate systems are determined by $n(n-1)/2$
functions of two variables (L.Bianchi). A striking fact can be discovered:
formula (3) was found in [6] (p. 353)! This formula is crucial for
the integrability property of diagonal hamiltonian systems (1): if we
interpret it as an overdetermined (compatible in view of zero curvature
property of $g$) system on $n$ unknown functions $v_{j}(u)$  ($g_{ii}$
given) we can generate from every its solution $\bar{v}_{j}(u) $ a
symmetry (commuting flow) $$
  u^{i}_{t} = \bar{v}_{i}(u)u^{i}_{x} ,\;\;  i = 1,...,n,
$$
of (1) and a solution of (1) (the generalized
hodograph method, see [33] for the details). One can prove ([33]) the
completeness property for this class of symmetries and solutions
parameterized by $n$ functions of 1 variable - the generic Cauchy data
for our diagonal system (1).

  The corresponding geometric notion used in the theory of orthogonal
curvilinear coordinate systems corresponding to (3) is the so called
Combescure transformation (see [6]).

 \underline{Definition}. Two orthogonal curvilinear coordinate systems
$x^i=x^i(u^1,\ldots ,u^n)$ and $\hat{x}^i=\hat{x}^i(u^1,\ldots ,u^n)$ in
the same flat (pseudo)Euclidean space $R^n =\{(x^1,\ldots ,x^n)\}$  are
said to be related by a Combescure transformation (or simply parallel) iff
their tangent frames $\vec{e}_i=\partial \vec{x}/\partial u^i $ and
$\hat{\vec{e}}_i=\partial \hat{\vec{x}}/\partial u^i $ are {\it parallel}
in points corresponding to the same values of  curvilinear coordinates
$u^i$.

  Let us take the quantities $ H_i(u)=|\vec{e}_i|=\sqrt{g_{ii}}$,
$\hat{H}_i(u)=|\hat{\vec{e}}_i| $  (Lam\'e coefficients).

 \underline{Proposition} The quantities  $\bar{v}_{i}(u)=\hat{H}_i(u)/H_i(u)$
satisfy (3) with $\Gamma ^{k}_{ki}=
    \partial_{i}H_k/H_k$, the connection coefficients for the metric
$g_{ii}=H_i^2$.  Conversely, for any solution $\bar{v}_{i}$  of (3)
$\hat{g}_{ii}=(\bar{v}_{i}H_i)^2$ will give an orthogonal curvilinear
coordinate system related to the coordinate system with the metric
$g_{ii}=H_i^2$ by a Combescure transformation.

  The theory of Combescure transformations coincides with the theory
of integrable diagonal systems of hydrodynamic type.

  Physical examples of such systems ( Whitham equations, Benney equations)
have hamiltonian structures (2) with diagonal metrics $g_{ii}$  possessing
the so called Egorov property: $\partial_{i}g_{kk}= \partial_{k}g_{ii}$.
 As we have demonstrated earlier ([34]) this
is a consequence of Galilei invariance of the original systems. See also
[12] for the algebro-geometric background of this property for averaged
integrable systems. Using this property and homogeneity of coefficients
one can find explicit formulas for solutions of (3) for the systems in
question [33], [34].

  The class of Egorov orthogonal curvilinear coordinate systems is
interesting in itself and merits our special attention.
\vspace{2em}

\noindent {\bf 2. Egorov coordinate systems, the $N$-wave problem and its
generalizations}
\vspace{2em}

\noindent Introducing $\beta _{ik}(u) = \partial _{i}H_{k}/H_{i},
\quad i\neq k$, $\beta _{ii}(u) = 0$ (rotation coefficients of a given
orthogonal curvilinear coordinate system with $g_{ii}=H_i^2$  , see [6])
one can easily check the following:

a) vanishing of the curvature tensor is equivalent to
\beq
\partial _{j}\beta _{ik}
= \beta _{ij}\beta _{jk} ,\qquad i\neq j\neq k,
\eeq
\beq
\partial _i\beta _{ik}+\partial _k\beta _{ki}+\sum_{s \neq i,k}
 \beta _{si}\beta _{sk}=0,\qquad  i\neq k.
\eeq
b) the Egorov property $\partial_{i}g_{kk}= \partial_{k}g_{ii}$
  reduces to
\beq
\beta _{ik}=\beta _{ki}.
\eeq
In the Egorov case condition (6) is equivalent to
$\hat{T} \beta _{ik} = 0, \hat T = \partial_{1} + \ldots +
\partial_{n}$.
Consequently the problem of classification of Egorov coordinate systems is
reduced to description of all off-diagonal symmetric matrices $ (\beta _{ik}) $
satisfying (5) and $\hat{T} \beta _{ik} = 0$.

  B.A.Dubrovin [12] have recently observed that this problem coincides with
the purely imaginary reduction of the well-known integrable system
describing resonant $N$-wave interactions. Namely, restriction of
 $\beta _{ik} $ on any $(x,t)$ plane $u^i=a^ix+b^it $ gives (compare, for
example, [28])
$$
[A,\Gamma _t]-[B,\Gamma _x]=[[A,\Gamma ],[B,\Gamma ]],
$$
$ A=diag(a^1, \ldots ,a^n),  B=diag(b^1, \ldots ,b^n),
\Gamma = (\beta _{ik}) $
 with additional reduction Im~$\Gamma =0, \Gamma ^T =\Gamma $.  For the
case $N=3$ this reduces to
\beq
 \left\{\bear{ccc} b_t^1 +c_1b_x^1 & = & \kappa b^2b^3 ,\\
   b_t^2 +c_2b_x^2 & = & \kappa b^1b^3 ,\\
   b_t^3 +c_3b_x^3 & = & \kappa b^1b^2 .
\ear \right .
\eeq
 This is a system of type (1+1,nh),integrable by the IST method ([28]).

  Now we can compare the progress achieved in the modern integrability
theory for (8) and the results obtained more than 70 years ago in the theory
of Egorov coordinate systems initiated by G.Darboux in 1866 and continued
by D.Th.Egorov in 1901 in his thesis (see [13]). It was Darboux [6] who
proposed to call this special type of coordinate systems Egorov type
systems. From the point of view of integrability properties remarkable
progress was achieved by L.Bianchi in 1915 (see [2]). He found a B\"acklund
transformation for
this problem and established the permutability property as well as the
superposition formula for it. We shall remark here that the pioneering
results on B\"acklund transformations
and their permutability in the well-known theory of
constant curvature surfaces in $R^3$ are due to Bianchi also.

 Let us take an orthogonal curvilinear coordinate system (not necessary of
Egorov type) with Lam\'e coefficients $H_i(u)$ and rotation coefficients
$\beta _{ik}(u)$.
Bianchi applied to it a generalization of Ribaucour transformations known
in the theory of transformations of surfaces. We recall that two surfaces
$\vec{x}(u,v)  $ and $\hat{\vec{x}}(u,v)   $ in $R^3$ are related by a
Ribaucour transformation  iff there exists a two-parametric family of
spheres $S(u,v) $ such that each sphere $S(u_0,v_0)$ is tangent to both
surfaces in corresponding points $\vec{x}(u_0,v_0),\hat{\vec{x}}(u_0,v_0)
$ {\it and} this correspondence preserves the curvature lines on the
surfaces. For the case of a pair of orthogonal curvilinear coordinate
systems in $R^3$ we need a three-parametric family of spheres (or an
$n$-parametric family for the $n$-dimensional case) tangent in the
corresponding points to one of three families  of coordinate surfaces as
well as to a coordinate surface of the other curvilinear coordinate
system. Since due to the classical Dupin theorem coordinate lines in any
orthogonal curvilinear coordinate system are curvature lines their
correspondence is guaranteed. In terms of the rotation coefficients
$\beta _{ik}$ one shall find a solution $\gamma _i(u)$ of
\begin{equation}
    \partial _{i}\gamma _{k} = \beta _{ki}\gamma _{i} ,\;\;   i\neq k,
\end{equation}
 to define the corresponding
Ribaucour transformation  ([2])
\beq
\hat{\beta }_{ik}=\beta _{ik}-\frac{2\gamma _i}{A}(\partial _k\gamma _k
 + \sum_{s\neq k}\beta _{sk}\gamma _s) ,\qquad A=\sum_p (\gamma _p)^2.
\eeq

For the case of Egorov systems we
shall complete (9) and restrict $\gamma _i$ to satisfy
$$
\partial _i \gamma _i = c\gamma _i -  \sum_{s\neq i}\beta _{si}\gamma _s,
\qquad (c=const)
$$
 or
\beq
\hat{T}\gamma _i = c\gamma _i,  \qquad \hat T = \partial_{1} + \ldots +
\partial_{n}
\eeq
Then the B\"acklund transformation  in question is
\beq
\hat{\beta }_{ik}=\beta _{ik}-\frac{2c\gamma _i \gamma _k}{A}.
\eeq

The permutability property for any orthogonal coordinate system requires a
quadrature, but for Egorov systems it may be found explicitly and
provides the following formulas for the fourth Egorov system
$\bar{\beta }_{ik}  $ related
to $\beta _{ik}', \beta _{ik}'' $ obtained from a given Egorov system
$\beta _{ik} $ with constants $c',c'' $ ($c'+c''\neq 0$) and potentials
$\gamma _i', \gamma _i''$ in (10):
$$
\bar{\gamma }_i'= \gamma _i'' - \frac{2c'\gamma _i
  \sum_s (\gamma _s'\gamma _s'')}{(c'+c'')\sum_s (\gamma _s')^2},
\bar{\gamma }_i''=  \gamma _i' - \frac{2c''\gamma _i
  \sum_s (\gamma _s'\gamma _s'')}{(c'+c'')\sum_s (\gamma _s'')^2}.
$$
   One can enjoy reading [4], [23] where these formulas were rediscovered in
the context of 3-wave system. So the basic integrability results for (8)
were established long ago by Darboux, Egorov and Bianchi certainly with
the exception of the IST transformation.

  An unexpected result (hidden in [6]) consists in existence of a {\it
homogeneous} system (1+1,h) of three equations related to (8) by a
nonlocal transformation. Geometrically this is trivial. Given an
orthogonal curvilinear coordinate system in $R^3$ we have in each its
point $P(x_0,y_0,z_0) $ the orthogonal 3-frame of tangent planes
\beq
z=p_k(x-x_0)+q_k(y-y_0)+z_0, \qquad k=1,2,3.
\eeq
Let us parameterize it by 3 functions
$A(x,y,z),B(x,y,z),C(x,y,z)$,coefficients $p_k,q_k $ of tangent planes being
three solutions of
$$
\left\{\bear{l} pq+Ap+Bq=0,\\ p^2-q^2+2(Cp+Hq)=0, \qquad 2(BC-AH)+1=0,\ear
 \right .
$$
different from the trivial solution $p=q=0$ ([3]).
Then the Frobenius compatibility conditions for these three families of
distributions (13) give a system of three homogeneous first-order equations
of type (2+1,h):
\beq
\left\{\bear{l} 2(AC_z-CA_z)=2C_y+B_y-A_x \\
  2(BH_z-HB_z)=2H_x+B_y-A_x  \\
  AH_z-HA_z+BC_z-CB_z=A_y-B_x   \ear \right .
\eeq
where $H=(2BC-1)/2A $ .

  For this system one can reformulate the B\"acklund-like transformation (10)
given in terms of $\beta _{ik}(u)$. A number of different transformations
producing (with quadratures) solutions of (14) parameterized by arbitrary
many functions of one variable may be found in [6]. Thus (14) is integrable
in a sense to be discussed elsewhere.

  If we will search for solutions of (14) which do not depend on $z$ then a
remarkable integrable (1+1,h) system of three equation appears. Since one
can easily prove the equivalence of $z$-independence in (14) and the Egorov
property (7) we have received a homogeneous system related to (8) by a
nonlocal change of variables. In Euler $\varphi ,\psi ,\theta  $
parameterization of orthogonal 3-frames it reads
\beq
 \left(\bear{c}\psi _t \\\theta _t \\ \varphi _t \ear \right)=
   \left(\bear{ccc} -\cos ^2\varphi  &
    -\sin \varphi \cos \varphi /\sin \theta  & 0 \\
  -\sin \theta \sin \varphi \cos \varphi & -\sin ^2 \varphi & 0 \\
 -\cos \theta (1+\cos ^2\varphi ) &
  -\sin \varphi \cos \varphi \cos \theta /\sin \theta  & 1 \ear \right)
   \left(\bear{c}\psi _x \\ \theta _x \\ \varphi _x \ear \right)
\eeq
This nonlocal change does not affect the existence of higher order
conserved densities. Recently Ferapontov [17] have proved the uniqueness
result for such 3$\times $3 homogeneous systems possessing higher-order
conserved densities: they may be transformed to (15) by reciprocal and
point transformations. Also another nonlocal transition from (8) to (15)
as given there.

   The matrix of (15) has constant eigenvalues $-1,0,+1$ but its
eigenvector fields (properly normalized) form $so(3)$ Lie algebra,
consequently (15) is a non-diagonalizable (1+1,h) integrable system.

  The complete system (14) certainly may be called a (2+1)-dimensional
generalization of the (1+1)-dimensional 3-wave system (8). Orthogonal
curvilinear coordinate systems in $R^n$ provide also only
a {\it (2+1)-dimensional} generalization of the (1+1)-dimensional $N$-wave
system since they are parameterized by $n(n-1)/2$ functions of {\it two}
variables (L.Bianchi).
\vspace{2em}

\noindent {\bf 3. Semihamiltonian diagonal systems and coordinate systems
with conjugate lines }

\vspace{2em}
\noindent The class of integrable diagonal systems (1) is wider than the
class of {\it hamiltonian} systems of this type. Namely, the property (4)
which is a weaker consequence of the hamiltonian property is sufficient
([33]).  Let us call a diagonal system {\em semihamiltonian}  if $n=2$ or
if $n>2$ and $v_{i}(u)$ satisfy (4). As a physical example of a
semihamiltonian (but non-hamiltonian for $n>3$) system one can mention the
ideal Langmuir chromatography and electrophoresis systems ([33]).

  To every semihamiltonian system we can relate a diagonal metric
 $g_{ii}$ via
$$\partial_{i}\ln\,g_{kk}/2 =\partial_{i}v_{k}/(v_{i}-v_{k})$$.
This metric is {\it not flat} in general though some
coefficients of the curvature tensor vanish as a consequence of (4). Namely
introducing  $ H _{i}= \sqrt{g_{ii}}$ ,\   $
\beta _{ik}(u) = \partial _{i}H_{k}/H_{i}$, we can find out that (4) is
equivalent to the set  (5) of equations on $\beta _{ik}$. Solutions of (5)
may be parameterized by $n(n-1)$ functions of 2 variables. This system
coincides with the compatibility conditions for a linear system
$$
    \partial _{i}\psi _{k} = \beta _{ik}\psi _{i} ,\;\;   i\neq k.
$$
Restricting (5) on 3-dimensional planes $u^i=a^ix+b^iy+c^iz$ in $R^n$ we
obtain (for general nonvanishing constants $a^i,b^i,c^i $) a (2+1,nh)
system on $n(n-1)$ quantities $ \beta _{ik}(x,y,z)$.

  As we have seen earlier the theory of hamiltonian diagonal systems (1) is
closely related to the theory of orthogonal curvilinear coordinate systems
in $R^n$. The geometric background for the theory of semihamiltonian
systems is given by the theory of coordinate systems with conjugate
coordinate lines (see [6] and [7], t. 4, ch. 12). A general (non-orthogonal)
coordinate system $\vec{x}(u_1,u_2,u_3)$ in $R^3$  is called a system with
conjugate coordinate lines (or simply a conjugate coordinate system) if on
every coordinate surface $S_{i_0}=\{u^i_0=const\} $ in every
point $P(x_0,y_0,z_0)$ the lines of
intersection of this surface with two other coordinate  surfaces belonging
to other one-parametric families of coordinate surfaces and containing
$P(x_0,y_0,z_0)$ are conjugate on $S_{i_0} $ (with respect to its second
fundamental form).  Every orthogonal curvilinear coordinate system is
conjugate due to Dupin theorem mentioned above. The theory of conjugate
coordinate systems was developed by Darboux and others and borrowed a lot
of results from the classical theory of conjugate coordinate nets on
surfaces in $R^3$ (known as "nets" or "r\'eseaux", see [14], [21],
[22], [36]).  A number of B\"acklund-like transformations for these
coordinate systems was given with permutability properties (though the
superposition formulas therein require quadratures).

  Every conjugate coordinate system  $ x^i=x^i(u^1,\ldots ,u^n)$
 in $R^n=\{(x^1,\ldots ,x^n) $ is
characterized by the conditions of conjugacy of coordinate lines:
\beq
\partial _i\partial _k\vec{x} = \Gamma ^k_{ki}(u)\partial _k\vec{x}
 +\Gamma ^i_{ik}(u)\partial _i\vec{x},\qquad i \neq k.
\eeq
This system of equations coincides with the system describing hydrodynamic
type conserved quantities of a semihamiltonian system with $    $  ([33]).
Quantities $\Gamma ^k_{ki}$ in (16) satisfy its compatibility conditions
$$
\partial _j\Gamma ^k_{ki}=\partial _i\Gamma ^k_{kj},\quad
\partial _j\Gamma ^k_{ki}=\Gamma ^k_{kj}\Gamma ^j_{ji}+
\Gamma ^k_{ki}\Gamma ^i_{ij}-\Gamma ^k_{ki}\Gamma ^k_{kj}, \quad
i \neq j \neq k.
$$
equivalent to the semihamiltonian property (4). Introducing $H_i(u)$ as
solutions of $\partial _iH_k(u)=\Gamma ^k_{ki}H_k(u)$
and $\beta _{ik}=\partial _{i}H_{k}/H_{i} ,\;\;   i\neq k,  $ we receive a
set of $\beta _{ik}$ satisfying (5). The converse is also true: given a
solution $\beta _{ik} $ of (5) one can find (a number of)semihamiltonian
systems related to it. Any semihamiltonian system also may be related to a
Combescure transformation of conjugate coordinate systems ([6]).

  This geometric interpretation provides another example of integrable
(2+1,h) system. Namely, given a conjugate coordinate system in $R^3$ one
can take the field of its (non-orthogonal) tangent 3-frames
$(\vec{e}_1,\vec{e}_2,\vec{e}_3)$ and
parameterize it by 6 independent functions $e^i_k(x,y,z),i=1,2,k=1,2,3$
(the coefficients $e^3_k$ may be set to 1 due to normalization). Then the
Frobenius compatibility conditions give 3 homogeneous first-order PDE on
$e^i_k(x,y,z),i=1,2,k=1,2,3$.
 Another 3 equations are given by the conjugacy condition
$ \det((\vec{e}_i \cdot \nabla )\vec{e}_k,\vec{e}_i,\vec{e}_k)=0, i<k$.
This system of 6 equations is a homogeneous (2+1,h) system in question.
Its $z$-independent solutions satisfy a (1+1,h) system enjoying properties
analogous to those of (15): it has constant eigenvalues $-1,0,+1$ (all
doubly degenerate) and 6 linearly independent fields of eigenvectors
forming (if properly normalized) a nontrivial Lie algebra. This remarkable
system will be studied in subsequent publications.
\vspace{2em}

\noindent {\bf 4. Additional topics}
\vspace{2em}

\noindent Recently O.I.Mokhov and E.V.Ferapontov [27], [16] found a nonlocal
generalization of the hamiltonian formalism of hydrodynamic type (2).
V.E.Ferapontov  communicated to the author about the further
generalization resulting in the following beautiful theorem: any
semihamiltonian system (1) has a nonlocal hamiltonian structure with
a hydrodynamic hamiltonian and a hamiltonian operator with (possibly
infinitely many) nonlocal terms similar to those in [16].
Grinevich [19] derived a series of nonlocal symmetries for Whitham equations
as well as original KdV equations.

     Weakly nonlinear semihamiltonian systems (i.e. systems (6) with
$ \partial _{i}v_{i}=0$ without summation on $i$, such systems are also
called "linearly degenerate") were studied in [15], [30].The theory of
such systems is connected to the theory of $n$-webs on Euclidean plane,
Dupin cyclids and St\"{a}ckel metrics (E.V.Ferapontov).  Among the results
are: quasiperiodic behavior of their solutions ([30]), complete
description of such systems and complete sets of their hydrodynamic
symmetries ([15]). E.V.Ferapontov communicated to the author the following
fact: any $n$-phase ($n$-zone) quasiperiodic (or a $n$-soliton) solution
of the KdV equation can be represented with a solution of a weakly
nonlinear semihamiltonian system $R^{i}_{t} = (\sum_{k\neq i}
R^{k})R^{i}_{x} , i=1,\ldots,n$.  These results shall be compared with
Curro and Fusco's results [5] in the soliton-like interactions of Riemann
simple waves for some $2\times 2$ systems.

  IST-like methods were developed in [18], [24], [25] for some diagonal
hamiltonian systems of physical importance. Certainly this approach shall
be related to our geometric methods.

  In a series of preprints (see [9],[26] and references therein)
I.M.Krichever and B.A.Dubrovin exposed a remarkable link between the theory
of Egorov coordinate systems and Witten-Dijgraagh-Verlinder-Verlinder
equations for the correlation functions of topological conformal field
theories proving their integrability.

  {\it Acknowledgement}.   The author enjoys the occasion to thank
Dr. E.V.Ferapontov and Professor J.Gibbons  for many useful discussions and
hospitality in Imperial College, London, where a part of this paper was
written as well as Professor P. Clarkson and our colleagues in Exeter for
their hospitality during the Workshop.

\end{document}